\documentclass[12pt,preprint]{aastex}

\newcommand{\iris}{{\em IRIS}}
\newcommand{\sdo}{{\em SDO}}
\newcommand{\goes}{{\em GOES}}
\newcommand{\siiv}{Si {\sc iv}}
\newcommand{\cii}{C {\sc ii}}
\newcommand{\mgii}{Mg {\sc ii}}
\newcommand{\fexxi}{Fe {\sc xxi}}

\shorttitle{}
\shortauthors{}

\begin{document}

\title{Different Signatures of Chromospheric Evaporation \\in Two Solar Flares Observed with \iris}

\author{Y. Li$^{1,3}$, M. D. Ding$^{2}$, J. Hong$^{2}$, H. Li$^{1}$, W. Q. Gan$^{1}$}
\affil{$^1$Key Laboratory of Dark Matter and Space Astronomy, Purple Mountain Observatory, Chinese Academy of Sciences, Nanjing 210034, China; yingli@pmo.ac.cn}
\affil{$^2$School of Astronomy and Space Science, Nanjing University, Nanjing 210023, China}
\affil{$^3$CAS Key Laboratory of Solar Activity, National Astronomical Observatories, Beijing 100012, China}

\begin{abstract}
We present different signatures of chromospheric evaporation in two solar flares observed by the {\em Interface Region Imaging Spectrograph} ({\iris}). In the B1.6 flare on 2016 December 6 (SOL2016-12-06T10:40), the transition region \siiv\ line and the chromospheric \cii\ and \mgii\ lines show blueshifts with low velocities up to 20 km s$^{-1}$ at the flare loop footpoints in the rise phase, indicative of a gentle chromospheric evaporation. While in the C1.6 flare on 2015 December 19 (SOL2015-12-19T10:51), the \siiv, \cii, and \mgii\ lines exhibit redshifts with velocities from several to tens of km s$^{-1}$ at the footpoints, which might suggest an explosive chromospheric evaporation. Explosive evaporation has been observed in many flares that were captured by \iris; however, gentle evaporation, especially manifested as blueshifts in the cool \siiv, \cii, and \mgii\ lines, has scarcely been reported. Our results bring some new insights into chromospheric evaporation in the \iris\ era.
\end{abstract}


\noindent

\noindent

\noindent

\noindent

\noindent

\noindent

\section{Introduction}
 
Chromospheric evaporation refers to drastic mass motions as a result of rapid energy deposition (or plasma heating) in the dense chromospheric layer during solar flares. From a modeling perspective, there are two kinds of dynamic responses of the solar atmosphere to the local plasma heating mainly depending on the heating flux of the electron beam \citep{fish85a,fish85b,fish85c}. For electron beams with an energy flux $\leq$10$^{10}$ erg cm$^{-2}$ s$^{-1}$, the heated plasma expands upward at a relatively low velocity (tens of km s$^{-1}$) without significant downflows, which is called ``gentle evaporation". When the energy flux reaches $\geq$3$\times$10$^{10}$ erg cm$^{-2}$ s$^{-1}$, an overpressure causes not only upflows of hot plasma at hundreds of km s$^{-1}$ but also downflows of cooler material (or chromospheric condensation) at tens of km s$^{-1}$. This process is referred to as ``explosive evaporation". Both types of chromospheric evaporation are confirmed in observations (e.g., \citealt{mill06a,mill06b}).

Chromospheric evaporation can be well diagnosed through spectroscopic observations. For a gentle evaporation, the upgoing plasma causes blueshifts (or blueshifted components) in some chromospheric lines and/or spectral lines formed above the chromosphere. While in an explosive evaporation, the upward mass motions generate blueshifts (or blueshifted components) usually in hot coronal lines, and the downward moving plasma produces redshifts (or red-wing enhancements) in cool chromospheric or transition region lines. Blueshifts/redshifts due to chromospheric evaporation/condensation have been reported in a large number of studies using spectroscopic data from different instruments. Early observations from the Bent and Bragg Crystal Spectrometer (BCS) on board the {\em Solar Maximum Mission} and {\em Yohkoh}/BCS detected blueshifted components at a few hundreds of km s$^{-1}$ in the Ca {\sc xix} 3.18 \AA\ line (e.g., \citealt{dosc80,anto82,anto83,canf90,wuls94,ding96}). Blueshifted line profiles of Fe {\sc xix} 592.23 \AA\ with velocities from tens to hundreds of km s$^{-1}$ were observed by the Coronal Diagnostic Spectrometer (CDS) on board the {\em Solar and Heliospheric Observatory} (e.g., \citealt{bros03,teri03,bros04,harr05,delz06,teri06}). Using CDS data, some redshifts at tens of km s$^{-1}$ were observed in the chromospheric line of He {\sc i} 522.21 \AA\ or 584.33 \AA\ (e.g., \citealt{delz06,mill06a}) and the transition region O {\sc v} 629.73 \AA\ line \citep{czay99,kami05,delz06,mill06a}. Similar redshifts were also detected in the chromospheric He {\sc ii} 256.32 \AA\ line and the transition region O {\sc vi} 184.12 \AA\ line from the Extreme-ultraviolet (EUV) Imaging Spectrometer (EIS) on board {\em Hinode} \citep{mill09,wata10,ying11,dosc13,youn13,ying15}. With a good temperature coverage, EIS observed not only redshifts (or multiple redshifted components) but also blueshifts (or blueshifted components) indicative of explosive evaporation in major flares as well as in microflares  \citep{chen10,wata10,ying11,dosc13,youn13,ying15,bros16}. In particular, the velocities of blueshifts and redshifts were found to scale with the temperature \citep{mill09,chen10,wata10,youn13}. 
 
In recent years, high spatial (subarcsecond) and temporal (several seconds) resolution observations from the {\em Interface Region Imaging Spectrograph} ({\iris; \citealt{depo14}}) have revealed some new results for chromospheric evaporation. Entirely blueshifted \fexxi\ 1354.08 \AA\ line profiles indicating evaporation flows have been observed in many \iris\ flares \citep{tian14,batt15,bros15,grah15,lido15,ying15,poli15,poli16,sady15,sady16,youn15,dudi16,zhan16a,lido17,ying17,lido18}. This may suggest that the subarcsecond resolution of \iris\ is sufficient to resolve the evaporation flow \citep{bros15,ying15,tian15,poli16}. Redshifts (or red-wing enhancements) caused by chromospheric condensation were also detected in the chromospheric lines of \mgii\ 2803.52 \AA\ (or 2796.35 \AA\ and 2791.59 \AA) and \cii\ 1335.71 \AA\ (or 1334.53 \AA) and the transition region line of \siiv\ 1402.77 \AA\ (or 1393.76 \AA) (e.g., \citealt{tian14,bros15,ying15,grah15,kerr15,zhan16b,bros17,bros18,tian18}). It should be mentioned that the cool \siiv, \cii, and \mgii\ lines usually show redshifts at flare footpoints, which suggests an explosive evaporation, in particular, when blueshifts in the hot \fexxi\ line are also observed. Table \ref{tab-iris} gives the Doppler shift measurements related to chromospheric evaporation using \iris\ data from previous studies. It is seen that explosive evaporation is detected in nearly all these flares above C-class. By contrast, gentle evaporation, especially manifested as blueshifts in the cool \siiv, \cii, and \mgii\ lines, has rarely been reported.

\begin{table}
\begin{center}
\small
\caption{Doppler Shift Measurements Related to Chromospheric Evaporation Using \iris\ Data in Previous Studies}
\label{tab-iris}
\begin{tabular}{ccclll}
\tableline
\tableline
\multicolumn{1}{c}{Flare} & Date & Cadence  & Blueshifts          & Redshifts            & References \\
                              Class  &          &    (s)        & (Evaporation)    & (Condensation)   & \\
\tableline
C1.6	 & 2014-04-19 & 31 &  \fexxi & \siiv, \cii, \mgii & \cite{tian14} \\
C1.9 & 2014-11-19 & 9.5 &   -      & \siiv, \cii, \mgii & \cite{warr16} \\
C3.1 & 2014-03-15 & 16.5 & \fexxi & \siiv, \cii & \cite{bros17} \\
C3.1 & 2015-10-16 & 6 &  -      & \siiv & \cite{zhan16b} \\
C4.2 & 2015-10-16 & 9.4 & \fexxi & \siiv & \cite{zhan16a} \\
C6.5 & 2014-02-03 & 75 & \fexxi &   -    & \cite{poli15} \\
M1.0 & 2014-06-12 & 21 & \fexxi & \cii  & \cite{sady15,sady16} \\
M1.1 & 2014-09-06 & 9.5 & \fexxi & \siiv & \cite{tian15} \\
M1.6 & 2015-03-12 & 5.2 & \fexxi & \siiv, \mgii & \cite{tian18} \\
         &                    &       & \fexxi & \siiv & \cite{bran16} \\
M1.8 & 2014-02-13 & 42 &    -     & \mgii & \cite{kerr15} \\
M2.3 & 2014-11-09 & 37 & \fexxi & \siiv, \cii, \mgii & \cite{ying17} \\
M3.7 & 2017-09-09 & 9.4 & \fexxi & \siiv, \cii, \mgii & \cite{bros18} \\
M7.1 & 2014-10-27 & 16.2 & \fexxi & \siiv & \cite{lido17,lido18} \\
M7.3 & 2014-04-18 & 9.4 & \fexxi & \siiv, O {\sc iv}, C {\sc i} & \cite{bros15} \\
         &                    &      &     -    & \siiv & \cite{bran15} \\
X1.0 & 2014-03-29 & 75 & \fexxi & \siiv, \cii, \mgii & \cite{batt15} \\       
        &                     &      & \fexxi & \siiv, \cii, \mgii & \cite{ying15} \\
        &                     &      & \fexxi &    -                   & \cite{youn15} \\
        &                     &      &    -     & Fe {\sc ii}        & \cite{kowa17} \\
X1.6 & 2014-09-10 & 9.4 & \fexxi & \siiv & \cite{tian15} \\
        &                    &      & \fexxi & \mgii & \cite{grah15} \\
        &                    &      & \fexxi &    -    & \cite{dudi16} \\
        &                    &      & \fexxi & C {\sc i} & \cite{lido15} \\
X1.6 & 2014-10-22 & 131 & \fexxi & C {\sc i} & \cite{lido15} \\
        &                     &      & \fexxi & \siiv & \cite{leek17} \\
X2.0 & 2014-10-27 & 26 & \fexxi & \siiv & \cite{poli16} \\
\tableline
\end{tabular}
\end{center}
The hyphen ``-" means that the authors did not include any hot lines (i.e., \fexxi) or cool lines (such as \siiv) from \iris\ in their study. 
\normalsize
\end{table}

In this paper, we present observational signatures of explosive as well as gentle chromospheric evaporation in two flare events with \iris~spectroscopic data. To our best knowledge, this is the first time to report a gentle evaporation, especially manifested as blueshifts in the cool \siiv, \cii, and \mgii\ lines, from the high cadence observations of \iris.

\section{Instruments and Data Reduction}
\label{sec-ins}

The observational data we used here are mainly from \iris, whose spatial, temporal, and spectral (or velocity) resolution can be as high as $0.\!\!^{\prime\prime}33$, 2 s, and 1 km s$^{-1}$, respectively. \iris\ provides slit-jaw images (SJIs) in four different passbands (1330, 1400, 2796, and 2832 \AA) as well as spectra in the near-ultraviolet (NUV, 2783--2834 \AA) and far-ultraviolet (FUV, 1332--1358 \AA~and 1389--1407 \AA) wavelengths with either a raster scan or sit-and-stare mode. The NUV and FUV passbands contain some important spectral lines, such as \mgii\ h at 2803.52 \AA\ and \mgii\ k at 2796.35 \AA, \cii\ at 1334.53 \AA\ and 1335.71 \AA, \siiv\ at 1393.76 \AA\ and 1402.77 \AA, and \fexxi\ at 1354.08 \AA. We focus on the \siiv\ 1402.77 \AA\ (with a formation temperature of $\sim$10$^{4.8}$ K), \cii\ 1335.71 \AA\footnote{Note that the \cii\ 1335.71 \AA\ line is blended with another \cii\ 1335.66 \AA\ line \citep{judg03}. However, we still choose this line rather than the \cii\ 1334.53 \AA\ line because the former one is stronger and thus has a better signal-to-noise ratio than the latter one, especially in small flares. The effect of the blending on line parameters is talked about below.} ($\sim$10$^{4.3}$ K), and \mgii\ k 2796.35 \AA\ ($\sim$10$^{4.0}$ K) lines in the present study. The \iris\ level 2 data are used here, which have been subtracted for dark current and corrected for flat field, geometrical distortion, and wavelength.

Regarding the spectral analysis, since the observed \siiv\ line profiles show a good Gaussian shape in general, we use a single Gaussian fitting to derive the total intensity, Doppler velocity, and width of this line. For the optically thick \cii\ and \mgii\ lines, a moment analysis is performed to the observed line profiles. When measuring the Doppler velocity, we use the average line center over the flaring region as the reference wavelength (because outside the flaring region for our events here, the \siiv\ and \cii\ lines have a very low signal-to-noise ratio). The obtained reference wavelengths are 1402.771 \AA, 1335.706 \AA, and 2796.357 \AA\ for \siiv, \cii, and \mgii, respectively, which are similar to the values used in some previous studies (e.g., 1402.775$\pm$0.020 \AA\ for \siiv\ in \citealt{jeff18}, and 1402.7847$\pm$0.0309 \AA\ for \siiv\ and 1335.7059$\pm$0.0267 \AA\ for \cii\ in \citealt{bros18}). The uncertainties in measured Doppler velocities are estimated to be about $\pm$6, $\pm$5\footnote{This includes the uncertainty coming from the \cii\ line blending, which is about $-$1 km s$^{-1}$.}, and $\pm$3 km s$^{-1}$ for the \siiv, \cii, and \mgii\ lines, respectively.

We also employ the EUV images at 131 \AA\ observed from the Atmospheric Imaging Assembly (AIA; \citealt{leme12}) on board the {\em Solar Dynamics Observatory} (\sdo), which represent the emission primarily from the hot plasma at a temperature of $\sim$11 MK during flares. AIA  observes EUV and UV images in multiple channels with a spatial resolution of $1.\!\!^{\prime\prime}2$ or $0.\!\!^{\prime\prime}6$ pixel$^{-1}$ and a cadence of 12 s or 24 s. Moreover, \goes\ provides the soft X-ray 1--8 \AA\ fluxes for the two flare events that are described in the following Section.

\section{Observations and Results}
\label{sec-res}

\subsection{The B1.6 Flare on 2016 December 6}

\subsubsection{Observation Overview}

\begin{figure*}
\centering
\includegraphics[width=16cm]{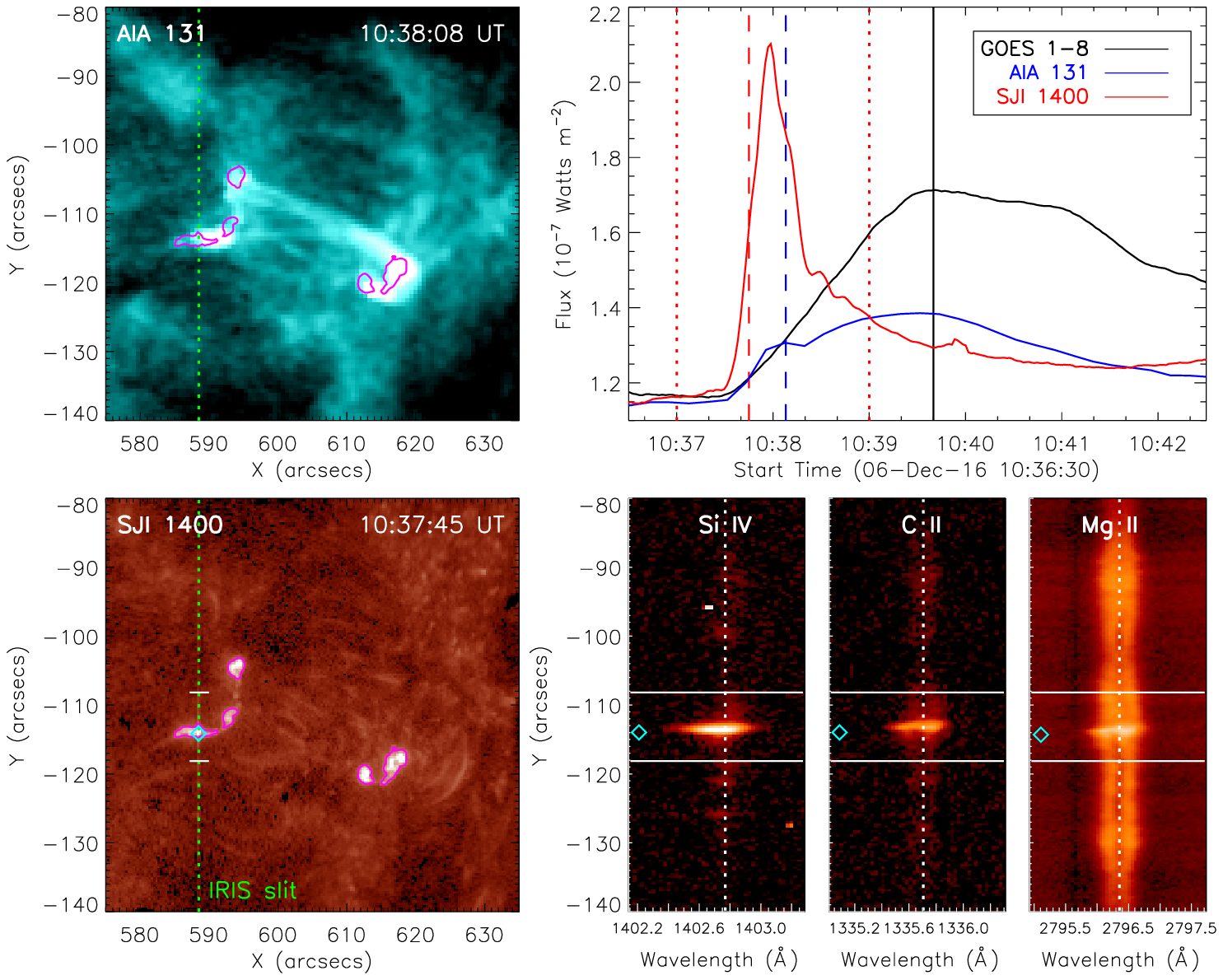}
\caption{{\small Observation overview of the B1.6 flare event on 2016 December 6. Top left panel: an AIA 131 \AA\ image during the rise phase of the flare. The green dotted line and the purple contours mark the \iris\ slit position and the flaring footpoints, respectively, which are the same as in the SJI 1400 \AA\ image (bottom left panel). Top right panel: light curves of \goes\ 1--8 \AA\ (black), AIA 131 \AA\ (blue, arbitrary scale), and SJI 1400 \AA\ (red, arbitrary scale) from the flaring region. The vertical black line represents the peak time of the flare. The blue and red dashed lines indicate the times when the AIA 131 \AA\ and SJI 1400 \AA\ images were observed, respectively. The two red dotted lines mark the time range (10:37:00--10:39:00 UT) as shown in Figures \ref{fig-mapb} and \ref{fig-evob}. Bottom right panel: spectra of \siiv, \cii, and \mgii\ along the slit at 10:37:45 UT. The vertical dotted line represents the reference wavelength for each of the spectral lines. The cyan diamond denotes the footpoint location that is cut across by the slit, as also shown in the SJI 1400 \AA\ image. The two horizontal lines mark the space range (10\arcsec) that is plotted in Figure \ref{fig-mapb}.}}
\label{fig-obsb}
\end{figure*}

The B1.6 microflare on 2016 December 6 has been reported by \cite{jeff18} on the topic of lower atmosphere turbulence deduced mainly from the \siiv\ line at flare ribbons, which will be discussed in Section \ref{sec-dis}. The observations of this flare is summarized in Figure \ref{fig-obsb}. Based on the \goes\ 1--8 \AA\ soft X-ray flux, the flare started at 10:37:00 UT, peaked at 10:39:40 UT, and ended around 10:43:10 UT, lasting for only several minutes. Using a sit-and-stare mode, \iris\ obtained the spectra of this flare for the whole period with an unprecedented high cadence of 1.7 s. SJIs were also recorded but only in 1400 \AA\ with a cadence of 2 s and a pixel scale of $0.\!\!^{\prime\prime}33$. From the AIA 131 \AA\ image during the rise phase of the flare, we can clearly see that the flare loops filled with hot ($\sim$11 MK) plasma map to some bright footpoints (the purple contours) in the SJI 1400 \AA\ passband. The \iris\ slit cuts across some of the footpoints (denoted by the cyan diamond) where the spectra of \siiv, \cii, and \mgii\ are fairly bright and show some blueshifts. 

\subsubsection{Blueshifts in the \siiv, \cii, and \mgii\ Lines}

\begin{figure*}
\centering
\includegraphics[width=16cm]{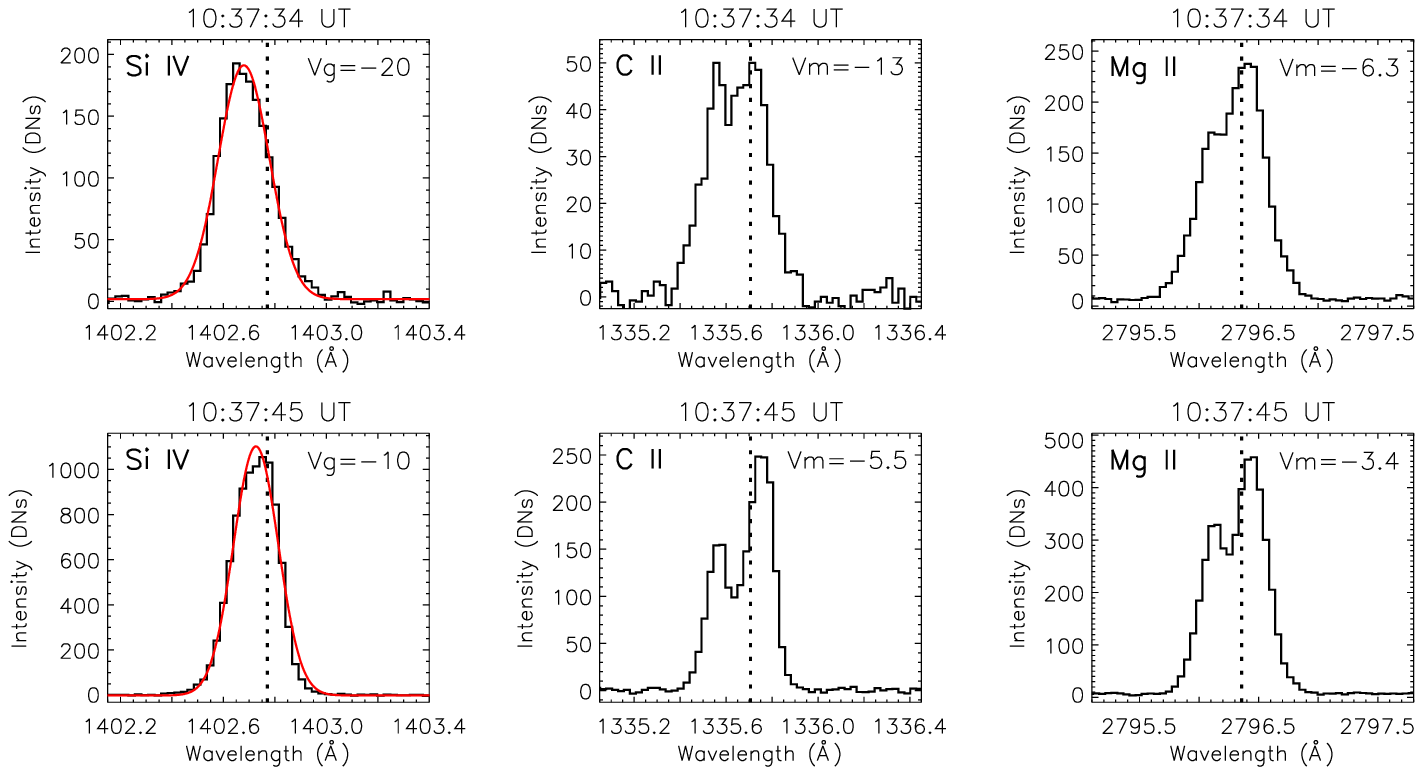}
\caption{{\small \siiv, \cii, and \mgii\ line profiles at the footpoint location denoted by the cyan diamond as shown in Figure \ref{fig-obsb} for two times. The dotted line in each panel represents the reference wavelength of the spectral line. For the \siiv\ line, we adopt a single Gaussian fitting (the red curve) to derive the Doppler velocity ($V_{g}$). For the optically thick \cii\ and \mgii\ lines, we make a moment analysis to obtain the Doppler velocity ($V_{m}$).}}
\label{fig-prob}
\end{figure*}

Figure \ref{fig-prob} shows the \siiv, \cii, and \mgii\ line profiles at the footpoint location denoted by the diamond (see Figure \ref{fig-obsb}) for two times. It is seen that all the three lines are blueshifted, especially at an earlier time (10:37:34 UT, top panels). The blueshift velocities are measured to be 20, 13, and 6.3 km s$^{-1}$ for the \siiv, \cii, and \mgii\ lines, respectively, namely decreasing as the line formation temperature decreases. A few seconds later (10:37:45 UT, bottom panels), all the velocities decrease nearly by half. Note that some of the \mgii\ (and also \cii) line profiles are almost singly peaked, i.e., without a central reversal, which is a common feature for flaring footpoints.

\begin{figure*}
\centering
\includegraphics[width=16cm]{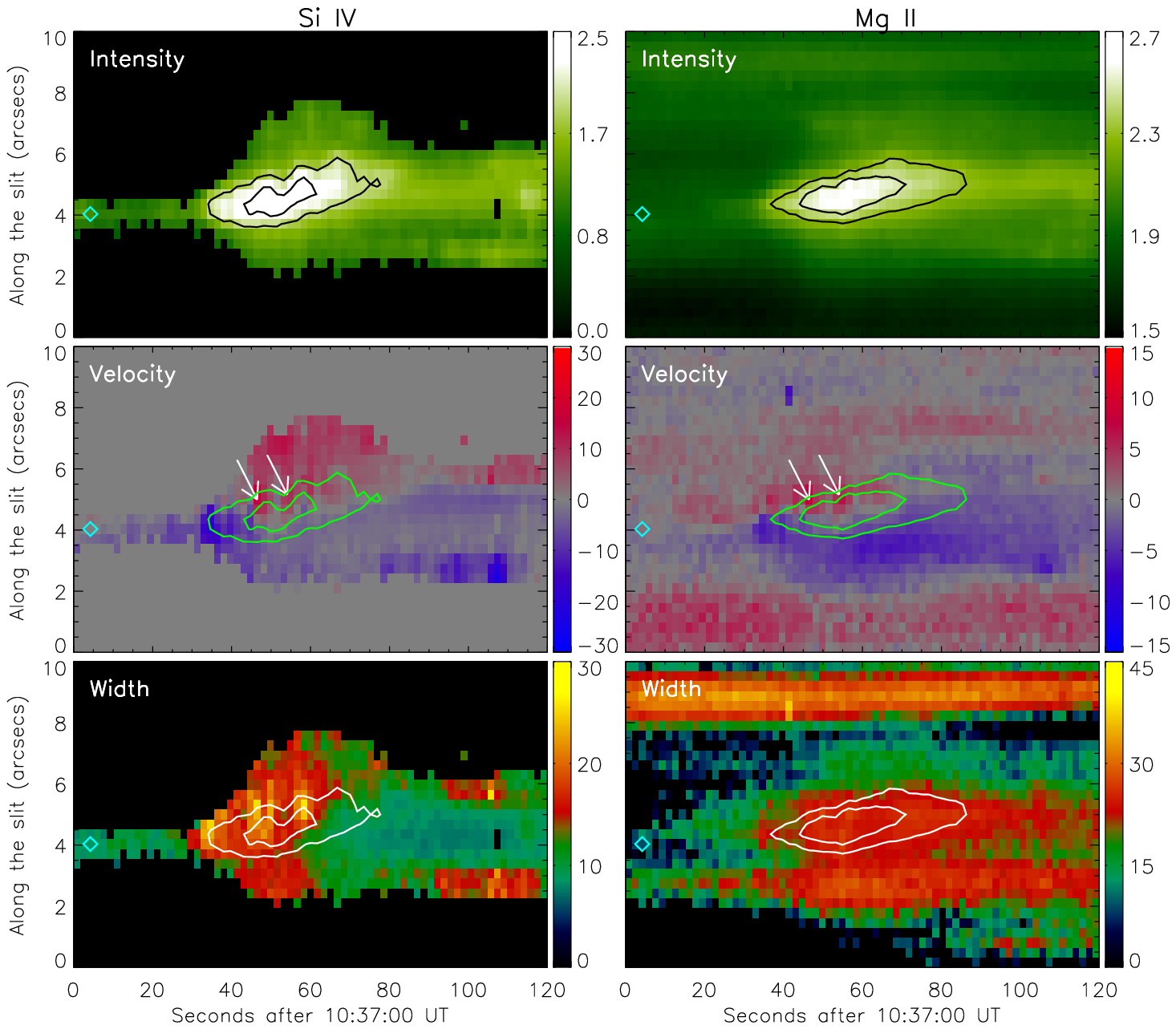}
\caption{{\small Spatio-temporal variations of the total intensity (in units of log (DNs)), Doppler velocity (in units of km s$^{-1}$), and line width (in units of km s$^{-1}$) for the \siiv\ and \mgii\ lines. The contours indicate the very bright footpoint region. The cyan diamond denotes the footpoint location as an example for study (the same as in Figure \ref{fig-obsb}). The white arrows mark the footpoint locations that show some redshifts.}}
\label{fig-mapb}
\end{figure*}

The spatio-temporal variations of the total intensity, Doppler velocity, and line width of the \siiv\ and \mgii\ lines are plotted in Figure \ref{fig-mapb}. From the intensity map we can see that the bright footpoints (the contours) show some apparent motion toward the north. Interestingly, at the footpoints (e.g., the ones denoted by the diamond), both of the \siiv\ and \mgii\ lines exhibit blueshifts particularly at an early time. Such a blueshift in cool lines has scarcely been reported before. A similar result is also obtained in the \cii\ line as can be seen in Figure \ref{fig-evob}. In addition, these blueshifts are accompanied by relatively large line widths. Note that some of the pixels also show redshifts and large line widths, which, however, are mostly located outside the core region of the bright footpoints.

Figure \ref{fig-evob} plots the temporal evolution of the total intensity, Doppler velocity, and line width at the diamond location (top panels) and its adjacent location (bottom panels). It is seen that, as the intensity increases, the \siiv\ and \mgii\ (and also \cii) lines begin to show some blueshifts with velocities up to 20 km s$^{-1}$, as well as large line widths, at both locations. The blueshift velocity reaches its maximum a few seconds earlier than the intensity. The deduced blueshifts above the uncertainty level in all of the three lines last for only about 10--20 s (from 10:37:24 UT to 10:37:45 UT, denoted by the two vertical dotted lines). Note that the \mgii\ line still exhibits notable blueshifts when its intensity decreases.

\subsubsection{Interpretation: Signatures of Gentle Chromospheric Evaporation}

The blueshifts, indicative of upflows, in the \siiv, \cii, and \mgii\ lines at the flare footpoints are consistent with the gentle evaporation scenario \citep{fish85a,mill06b}, in which the plasma in the upper chromosphere and above moves upward with a relatively low velocity. In this microflare, all the three lines formed in the chromosphere or transition region are blueshifted with velocities up to only $\sim$20 km s$^{-1}$ when the line intensity increases. After the intensity peak, only the cooler \mgii\ line continues showing blueshifts. A possible reason is that at a later time, when the heating rate becomes smaller, the evaporated plasma is not heated to a high temperature as visible in the \siiv\ and \cii\ lines. We also notice that all the blueshifts are accompanied by relatively large line widths, which is often an observational signature for the evaporation plasma \citep{ying15,ying17}. Note that, besides gentle evaporation, we also find evidence of explosive evaporation, i.e., redshifts of the lines at some specific footpoint locations (see the white arrows in Figure \ref{fig-mapb}), where the heating, reflected from the line intensity, seems to be stronger than the flaring regions that exhibit blueshifts.

\subsection{The C1.6 Flare on 2015 December 19}

\subsubsection{Observation Overview}

\begin{figure*}
\centering
\includegraphics[width=16cm]{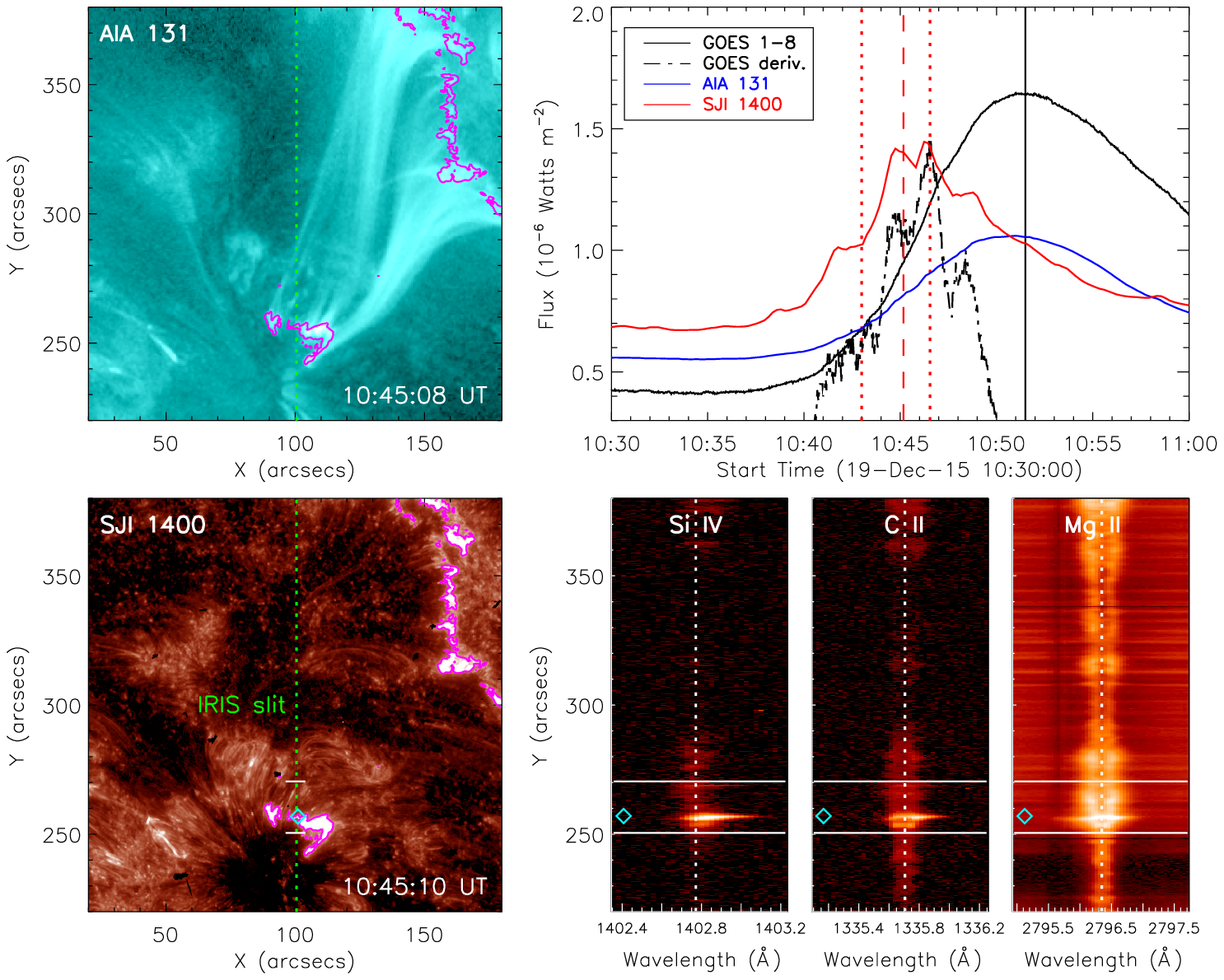}
\caption{{\small Observation overview of the C1.6 flare event on 2015 December 19. Top left panel: an AIA 131 \AA\ image during the rise phase of the flare. The green dotted line and the purple contours mark the \iris\ slit position and the flaring footpoints, respectively, which are the same as in the SJI 1400 \AA\ image (bottom left panel). Top right panel: light curves of \goes\ 1--8 \AA\ (black solid), AIA 131 \AA\ (blue solid, arbitrary scale), and SJI 1400 \AA\ (red solid, arbitrary scale) from the flaring region. Also shown is the time derivative of the \goes\ 1--8 \AA\ light curve (black dotted, arbitrary scale). The vertical black line represents the peak time of the flare. The red dashed line indicates the time when the SJI 1400 \AA\ as well as AIA 131 \AA\ images were observed. The two red dotted lines mark the time range (10:43:00--10:46:33 UT) as shown in Figures \ref{fig-mapc} and \ref{fig-evoc}. Bottom right panel: spectra of \siiv, \cii, and \mgii\ along the slit at 10:45:10 UT. The vertical dotted line represents the reference wavelength for each of the spectral lines. The cyan diamond denotes the footpoint location that is cut across by the slit, as also shown in the SJI 1400 \AA\ image. The two horizontal lines mark the space range (20\arcsec) that is plotted in Figure \ref{fig-mapc}.}}
\label{fig-obsc}
\end{figure*}

Figure \ref{fig-obsc} gives an observation overview of the C1.6 flare on 2015 December 19. This flare started at 10:40 UT and peaked around 10:51 UT based on the \goes\ 1--8 \AA\ soft X-ray flux. \iris\ captured the whole flare period with a sit-and-stare mode that yields spectral data with a very high cadence of 3 s, and also recorded SJIs in 1400, 1330, 2796, and 2832 \AA\ passbands with a cadence of 13 s and a pixel scale of $0.\!\!^{\prime\prime}33$. Here we only focus on part of the rise phase from 10:43 UT to 10:46 UT (denoted by the two vertical dotted lines in the top right panel of Figure \ref{fig-obsc}). After 10:47 UT, there is a severe contamination in the \iris\ data due to some compact cosmic rays. From the AIA 131 \AA\ image, we can see that some flare loops filled with hot ($\sim$11 MK) plasma show up in the rise phase with their footpoints (or flare ribbons) clearly visible in the SJI 1400 \AA\ passband (see the purple contours). The \iris\ slit cuts across one of the ribbons, as denoted by the cyan diamond, where the \siiv, \cii, and \mgii\ lines show significant emissions with some redshifts.

\subsubsection{Redshifts in the \siiv, \cii, and \mgii\ Lines}

\begin{figure*}
\centering
\includegraphics[width=16cm]{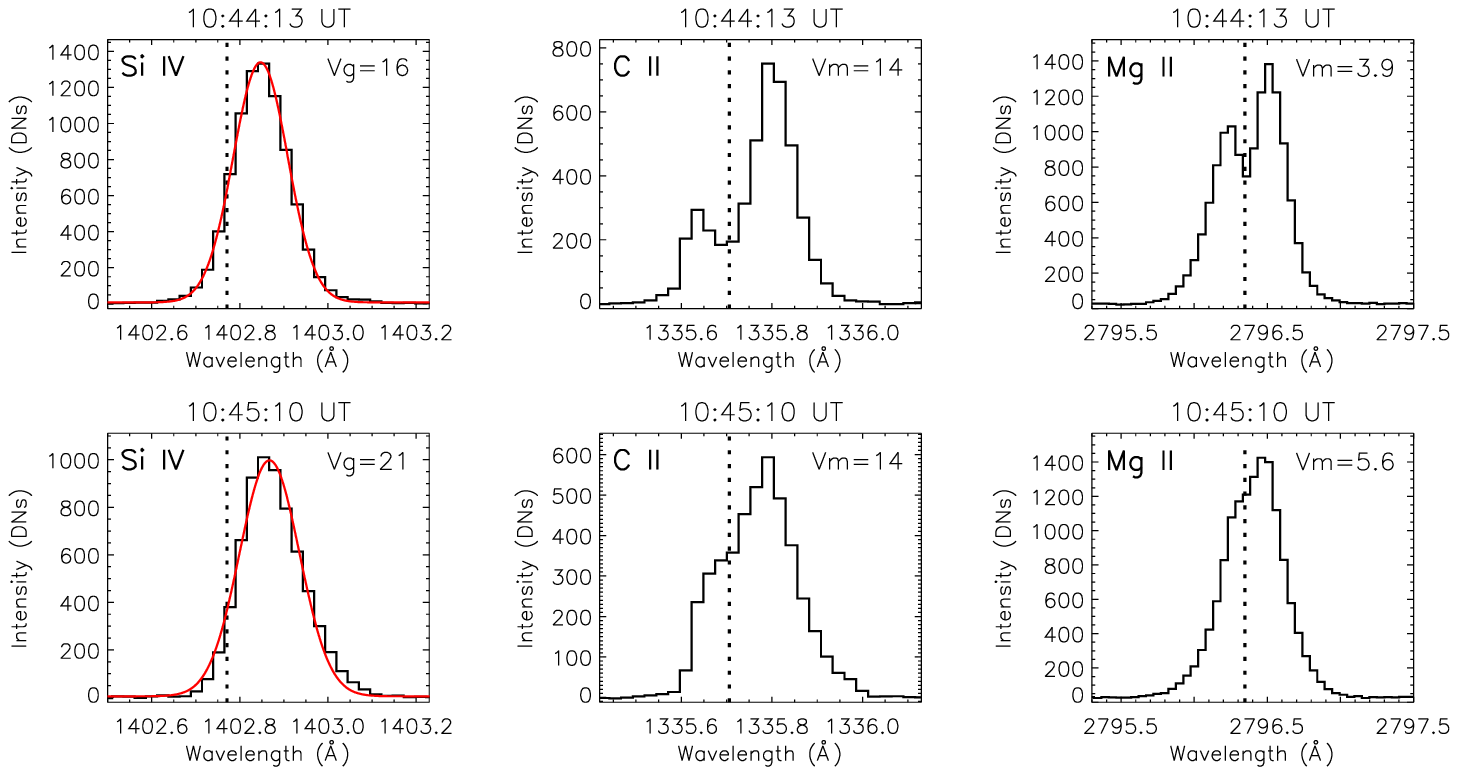}
\caption{{\small \siiv, \cii, and \mgii\ line profiles at the footpoint location denoted by the cyan diamond as shown in Figure \ref{fig-obsc} for two times. The dotted line in each panel represents the reference wavelength of the spectral line. For the \siiv\ line, we adopt a single Gaussian fitting (the red curve) to derive the Doppler velocity ($V_{g}$). For the optically thick \cii\ and \mgii\ lines, we make a moment analysis to obtain the Doppler velocity ($V_{m}$).}}
\label{fig-proc}
\end{figure*}

Figure \ref{fig-proc} shows the \siiv, \cii, and \mgii\ line profiles at the footpoint location denoted by the diamond (see Figure \ref{fig-obsc}) for two times. It is seen that all the three lines are redshifted or show a red wing enhancement. At an earlier time (10:44:13 UT, top panels), the redshift velocities are measured to be 16, 14, and 3.9 km s$^{-1}$ for the \siiv, \cii, and \mgii\ lines, respectively. The velocities again show a decreasing trend as the temperature decreases. About 1 minute later (10:45:10 UT, bottom panels), the redshift velocities of \siiv\ and \mgii\ increase to 21 and 5.6 km s$^{-1}$, respectively. Note that some of the \mgii\ and \cii\ line profiles are singly peaked without a central reversal.

\begin{figure*}
\centering
\includegraphics[width=16cm]{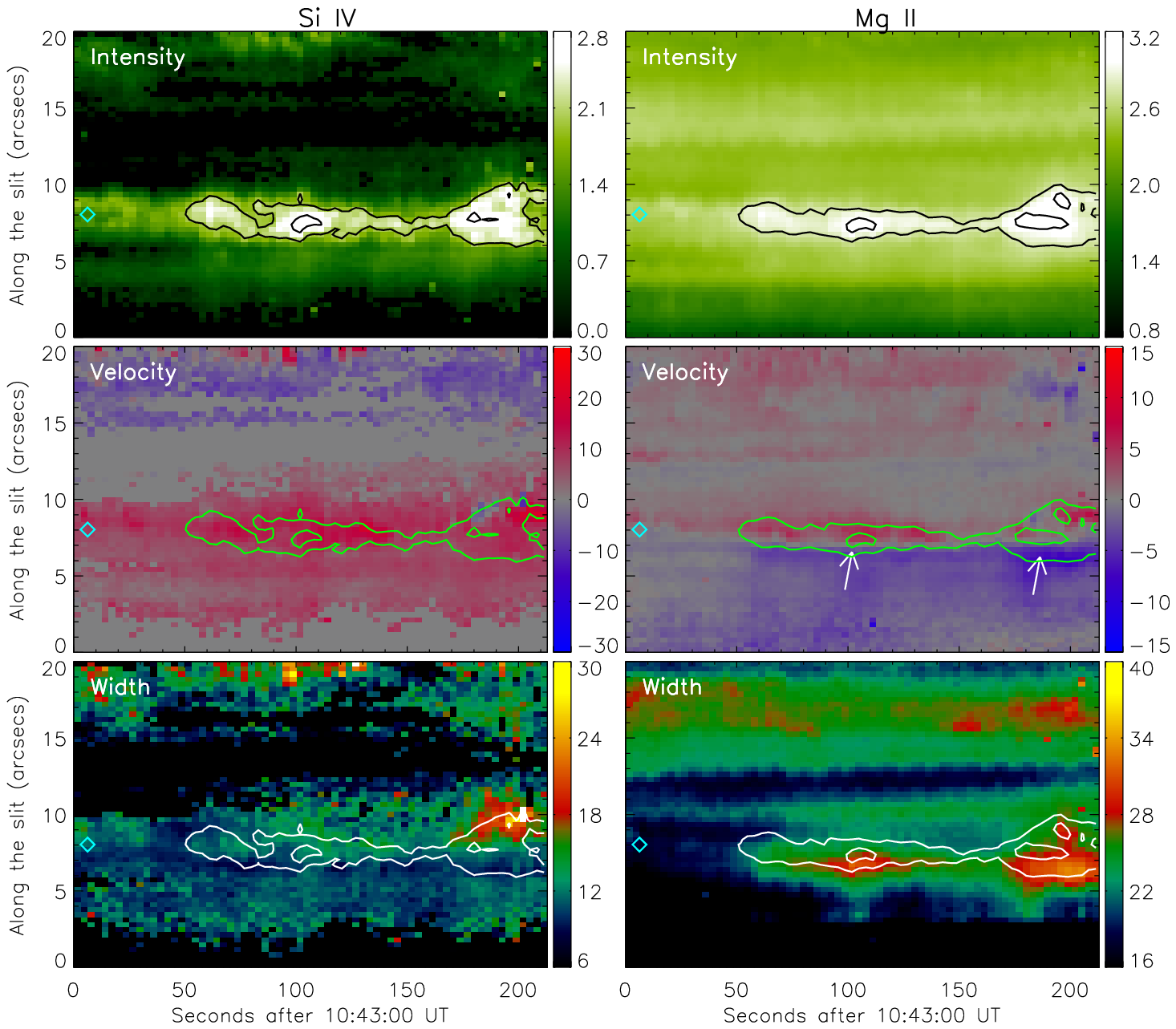}
\caption{{\small Spatio-temporal variations of the total intensity (in units of log (DNs)), Doppler velocity (in units of km s$^{-1}$), and line width (in units of km s$^{-1}$) for the \siiv\ and \mgii\ lines. The contours indicate the very bright footpoint region. The cyan diamond denotes the footpoint location as an example for study (the same as in Figure \ref{fig-obsc}). The white arrows mark the footpoint locations where the \mgii\ line show some blueshifts.}}
\label{fig-mapc}
\end{figure*}

\begin{figure*}
\centering
\includegraphics[width=16cm]{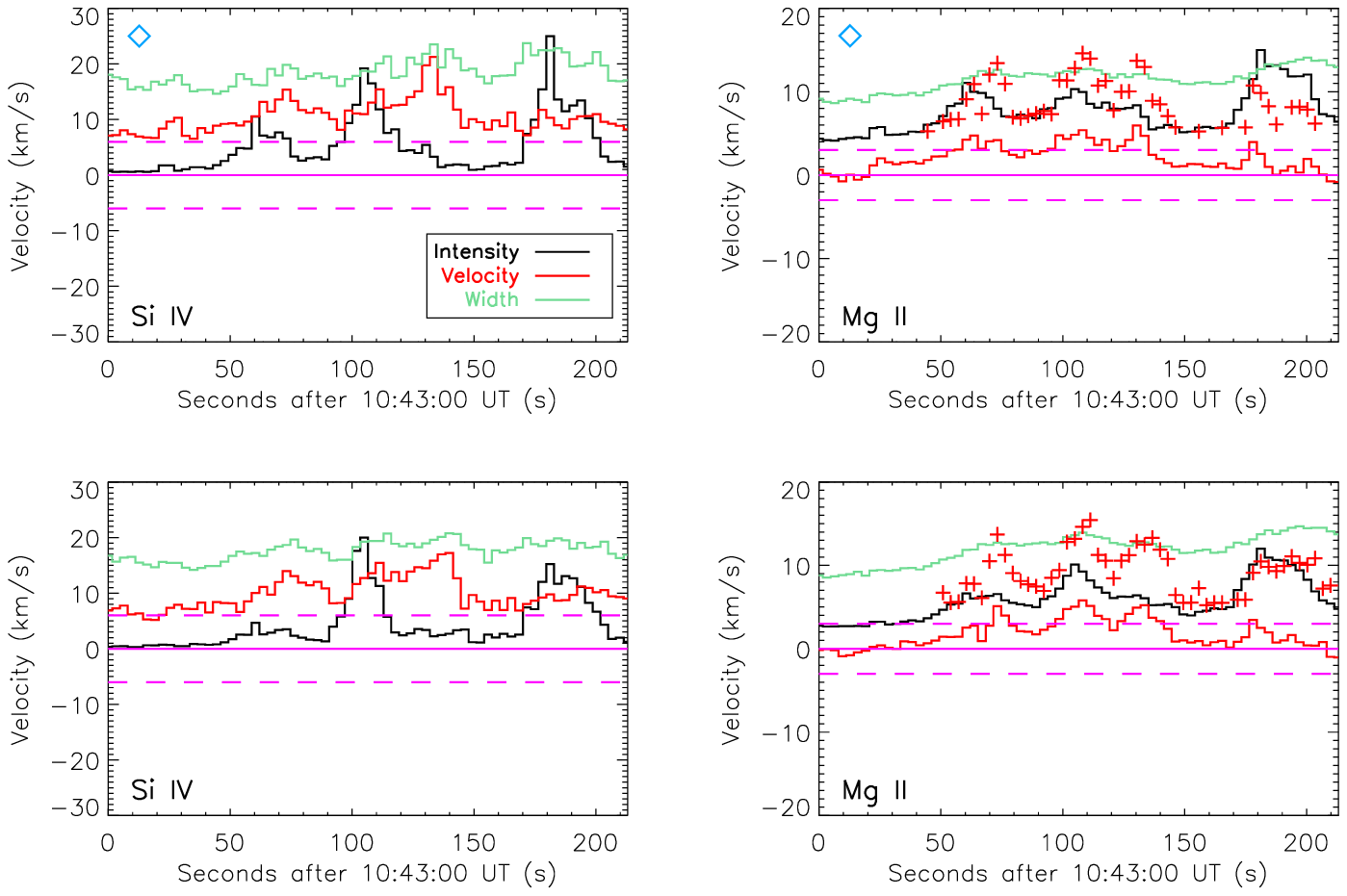}
\caption{{\small Temporal evolution of the total intensity (black, in an arbitrary scale), Doppler velocity (red), and line width (green) of the \siiv\ and \mgii\ lines at the diamond location (top panels) and a location adjacent to it (bottom panels). Note that the line widths of \siiv\ have been multiplied by 1.5 and the line widths of \mgii\ divided by 2 for clarity. The plus symbols in the \mgii\ panels represent the Doppler velocities of \cii, which are above the velocity uncertainties in measurements that are shown as the two horizontal dashed lines in each panel.}}
\label{fig-evoc}
\end{figure*}

The spatio-temporal variations of the total intensity, Doppler velocity, and line width of \siiv\ and \mgii\ are plotted in Figure \ref{fig-mapc}. It is seen that the ribbon crossed by the \iris\ slit stays there but brightens up repetitively. Around the footpoint location marked by the diamond, both of the \siiv\ and \mgii\ lines exhibit redshifts with a velocity of several tens of km s$^{-1}$, which has been reported in many studies before (see Table \ref{tab-iris}). A similar result is also shown in the \cii\ line as can be seen in Figure \ref{fig-evoc}. Note that, at some footpoints, the \mgii\ line displays weak blueshifts (denoted by the white arrows) while the \siiv\ line still shows redshifts, which will be discussed in Section \ref{sec-interpc}.

Figure \ref{fig-evoc} plots the temporal evolution of the total intensity, Doppler velocity, and line width at the diamond location (top panels) and its adjacent location (bottom panels). It is seen that, as the intensity increases, the \siiv\ and \mgii\ (and also \cii) lines begin to show some redshifts with velocities up to $\sim$20 km s$^{-1}$ at both locations. The intensity and Doppler velocity (and also line width) seem to exhibit an oscillation pattern, although these quantities do not reach the maximum values at the same time.

\subsubsection{Interpretation: Signatures of Explosive Chromospheric Evaporation}
\label{sec-interpc}

The redshifts, indicative of downflows, in the \siiv, \cii, and \mgii\ lines at the flare footpoints are supposed to be signatures of chromospheric condensation, which is usually associated with an explosive evaporation \citep{fish85a,mill06a}. In this scenario, the chromosphere is unable to fully radiate away the flare energy deposited there resulting in an overpressure which causes the local plasma going downward. In such a case, the cool lines are seen to be redshifted. We also check the hot emission line of \fexxi\ 1354.08 \AA\ ($\sim$10 MK), which is usually used to study the evaporation signatures. However, no visible \fexxi\ emission is found in this microflare. It seems that the flare is not strong enough to produce sufficient \fexxi\ photons within a short  exposure time of 2 s. Nevertheless, we do see some hot ($\sim$11 MK) plasma in AIA 131 \AA\ images (see the top left panel in Figure \ref{fig-obsc}), which fills the flare loop quickly after the footpoints brighten up. In fact, the time profile of the UV emission at the footpoints matches the derivative of the \goes\ 1--8 \AA\ flux very well (see the top right panel of Figure \ref{fig-obsc}). This relationship is similar to the Neupert effect \citep{neup68} that is often quoted as a manifestation of chromospheric evaporation driven by nonthermal electrons (e.g., \citealt{denn93,tian15}). In this flare, the redshifts of the \siiv, \cii, and \mgii\ lines display an oscillation pattern, which may imply a multi-episode explosive evaporation caused by episodic heating \citep{tian18}. There is also a possibility that some redshifts of the Si IV line, mainly those strong ones after the episodic intensity peaks, could be caused by cooling downflows \citep{tian18}. However, we think that the condensation scenario is mostly valid for this flare. Note that we also detect blueshifts in the cooler \mgii\ line (with a central reversal in the line core) but redshifts in the \siiv\ (and also \cii) lines at some footpoints. This is similar to the finding by \cite{teia18} who explained it as a result of cool plasma lifted up by evaporated hot plasma due to the penetration of nonthermal electrons into the chromosphere.

\section{Summary and Discussions}
\label{sec-dis}

In this paper, we present different signatures of chromospheric evaporation in two flare events observed by {\iris} with an unprecedented high cadence. In the B1.6 flare, the cool \siiv, \cii, and \mgii\ lines show blueshifts with the velocities up to 20 km s$^{-1}$ at the footpoints, consistent with a gentle evaporation scenario. While in the C1.6 flare, all the three lines exhibit redshifts with velocities from several to tens of km s$^{-1}$, implying an explosive evaporation. Explosive evaporation has been detected in lots of flares that were observed by \iris. However, this is the first time, to the best of our knowledge, to report a gentle evaporation in \iris\ flares, especially manifested as blueshifts appearing simultaneously in the \siiv, \cii, and \mgii\ lines.

\begin{figure*}
\centering
\includegraphics[width=16cm]{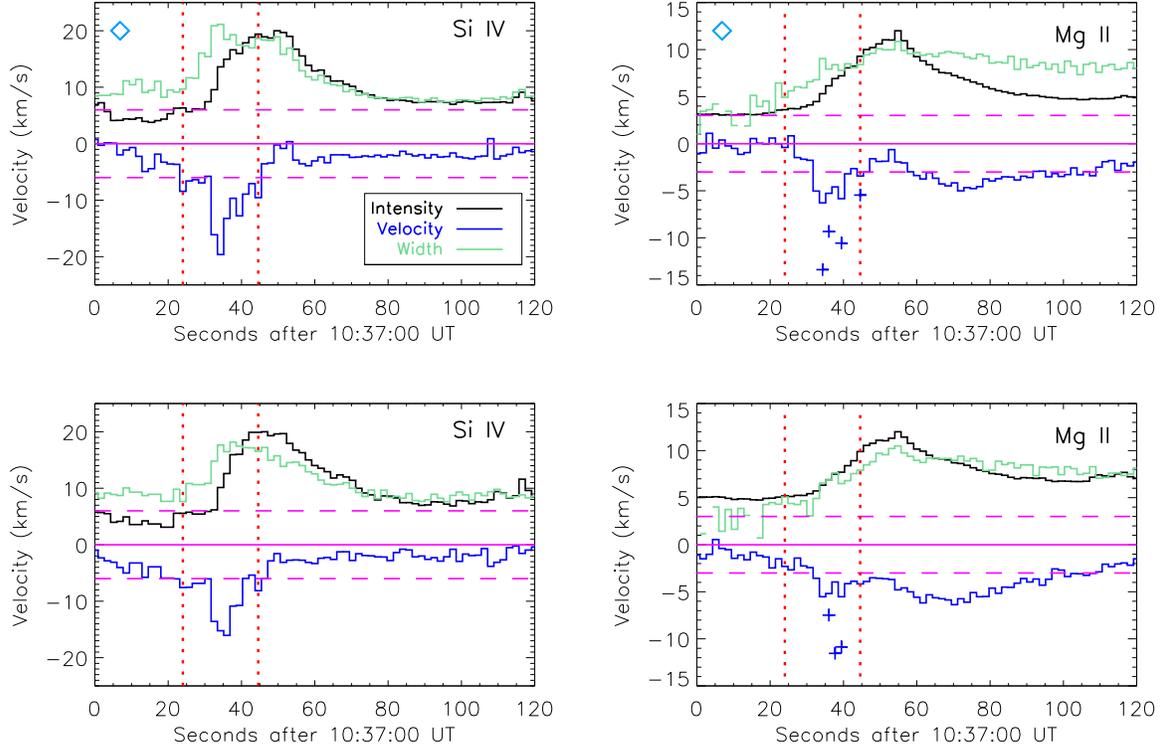}
\caption{{\small Temporal evolution of the total intensity (black, in an arbitrary scale), Doppler velocity (blue), and line width (green) of the \siiv\ and \mgii\ lines at the diamond location (top panels) and a location adjacent to it (bottom panels). Note that the values of the \mgii\ line widths have been divided by 2.5 for clarity. The plus symbols in the \mgii\ panels represent the Doppler velocities of \cii, which are above the velocity uncertainties in measurements that are shown as the two horizontal dashed lines. The two vertical dotted lines in each panel mark the time period (10:37:24--10:37:45 UT) when all the \siiv, \cii, and \mgii\ lines exhibit notable blueshifts.}}
\label{fig-evob}
\end{figure*}

Blueshifts (or bulk velocities) as well as large line widths (or line broadenings) in the \siiv\ line have been observed at the footpoints in the same B1.6 microflare by \cite{jeff18}. However, the authors focused on the line broadenings at the flare onset, which oscillate two or three times with a period of $\sim$10 s, and interpreted them as a result of turbulence in the lower solar atmosphere contributing to plasma heating. Here, we do find that the \siiv\ line widths (maybe also blueshifts) exhibit a weak fluctuation pattern at some footpoints, such as the diamond location as shown in the top left panel of Figure \ref{fig-evob}. However, such fluctuation behavior in line widths (or blueshifts) do not show at some other footpoints, for instance, the slit pixel adjacent to the diamond location (the bottom left panel of Figure \ref{fig-evob}). More importantly, the notable blueshifts in our study are well associated with enhanced intensities indicative of plasma heating. This is different from the case in \cite{jeff18} that the blueshifts as well as line broadenings appear well before the intensity rises.

Gentle evaporation has been observed in a few flares by EIS (e.g., \citealt{ying11}), CDS (e.g., \citealt{mill06b}), and some other instruments (e.g., \citealt{schm90}), though most of the studies were based on spectral data at only one time instant and at some of the footpoints. It has rarely been reported by using the \iris\ data in previous studies possibly for the two following reasons. First, the majority of the flares that were selected for study on chromospheric evaporation are large ones (at least two thirds are M-class or above as seen in Table \ref{tab-iris}), which are more likely to produce an explosive evaporation according to numerical simulations (e.g., \citealt{fish85a}). Second, for the weak flares (i.e., C-class), the cadence of \iris\ spectra ($\geqslant$6 s as seen in Table \ref{tab-iris}) seems not high enough to capture the short-lived gentle evaporation process (say, several seconds). Moreover, a gentle evaporation may only appear at a small part of footpoint pixels, which is not easy to be identified in most cases. In fact, in the microflare under study, there exist some footpoint pixels showing signatures of explosive evaporation, just like the footpoints of the C1.6 flare. In the future, it is worth checking more low-class flares observed by \iris\ with a high cadence (say, $\leqslant$5 s) to study the signatures of chromospheric evaporation, particularly combined with radiative hydrodynamic simulations.

Blueshifts in the \siiv\ line caused by chromospheric evaporation have actually been reproduced in radiative hydrodynamic simulations \citep{poli18,kerr19}. In particular, it is found that the lifetime of \siiv\ blueshifts becomes shorter when the heating flux is increased. For example, \cite{kerr19} showed that, assuming a constant heating rate for 10 s, the \siiv\ line exhibits blueshifts for more than 10 s in reletively weak heating cases (i.e., heating fluxes of $5\times10^{8}$, $1\times10^{9}$, and $5\times10^{9}$ erg cm$^{-2}$ s$^{-1}$), for $\sim$10 s in moderate heating cases (heating fluxes of $1\times10^{10}$ and $5\times10^{10}$ erg cm$^{-2}$ s$^{-1}$), and for only $\sim$5 s in a strong heating case (heating flux of $1\times10^{11}$ erg cm$^{-2}$ s$^{-1}$). In the latter one, the \siiv\ line can also show redshifts. These indicate that in small flares, the \siiv\ line blueshifts could be more likely observed than in large flares. This also confirms the general point of view that gentle evaporation more probably appears in small flares, while explosive evaporation more likely shows up in large flares. Our observational results are consistent with this picture.

Overall, the different observational signatures of chromospheric evaporation in these two flare events bring some new insights into chromospheric evaporation in the \iris\ era. In particular, we report blueshifts in the \siiv\ line that are rarely observed by \iris. Such line blueshifts can be well reproduced by the state-of-the-art radiative hydrodynamic simulations and thus provide observational support to the present flare heating models. We should also point out that, for the flares under study, the \siiv\ line profiles are well Gaussian shaped and are usually blueshifted or redshifted as a whole at the footpoints, while this line shows only a red asymmetry (a rest component plus a redshifted component) in many other flares (e.g., \citealt{ying15,tian15}). Such different spectral features have some implications on the flare heating and line formation processes that will be explored using radiative hydrodynamic simulations in the future.


\acknowledgments
{\em IRIS} is a NASA small explorer mission developed and operated by LMSAL with mission operations executed at NASA Ames Research center and major contributions to downlink communications funded by the Norwegian Space Center (NSC, Norway) through an ESA PRODEX contract. {\em SDO} is a mission of NASA's Living With a Star Program. The authors are supported by NSFC under grants 11873095, 11733003, 11961131002, 11427803, and U1731241, and by the CAS Strategic Pioneer Program on Space Science under grants XDA15052200, XDA15320301, and XDA15320103-03. Y.L. is also supported by the CAS Pioneer Hundred Talents Program, CAS Key Laboratory of Solar Activity of National Astronomical Observatories (KLSA201712), and ISSI-BJ from the team ``Diagnosing Heating Mechanisms in Solar Flares through Spectroscopic Observations of Flare Ribbons" led by Hui Tian.

\bibliographystyle{apj}

\end{document}